\documentclass[aps,twocolumn,floats,prl]{revtex4}
\usepackage{graphics,graphicx,epsfig}
\usepackage{amssymb}

\newcommand {\E}[1]{\cdot 10^{#1}}     	

\begin{document}

\title{Information flow and optimization in transcriptional control}

\author{Ga\v{s}per Tka\v{c}ik$,^{a}$ Curtis G. Callan, Jr.$,^{a,b}$ and William Bialek$^{a,b}$}

\affiliation{$^{a}$Joseph Henry Laboratories of Physics, $^a$Lewis--Sigler Institute for Integrative Genomics, and $^{b}$Princeton Center for Theoretical Physics,
Princeton University, Princeton, New Jersey 08544}

\begin{abstract}
In the simplest view of transcriptional regulation, the expression of a gene is turned on or off by changes in the concentration of a transcription factor (TF). We use recent data on noise levels in gene expression to show that it should be possible to transmit much more than just one regulatory bit. Realizing this optimal information capacity would require that the dynamic range of TF concentrations used by the cell, the input/output relation of the regulatory module, and the noise levels of binding and transcription satisfy certain matching relations.   This parameter--free prediction is in good agreement with recent experiments on the Bicoid/Hunchback system in the early {\em Drosophila} embryo, and this system achieves $\sim 90\%$ of its theoretical maximum information transmission.
\end{abstract}

\date{\today}
\maketitle 

\begin{figure}[b] 
   \centering
\includegraphics[width=\linewidth]{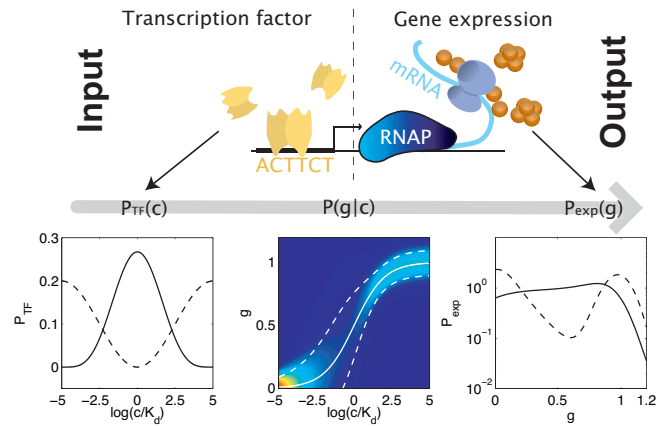}
      \caption{Transcriptional regulation of gene expression. The occupancy of the binding site by transcription factors sets the activity of the promoter and hence the amount of protein produced. The physics of TF--DNA interaction, transcription and translation processes determine the conditional distribution of expression levels $g$ at fixed TF concentration $c$, $P(g|c)$, shown here as a heat map with red (blue) corresponding to high (low) probability. The mean input/output relation is shown as a thick white line, and the dashed lines indicate $\pm$ one standard deviation of the noise around this mean.
Two sample input distributions $P_{TF}(c)$ (lower left) are passed through $P(g|c)$  to yield two corresponding distributions of outputs, $P_{\rm exp}(g)$ (lower right).}
\label{f-scheme}
\end{figure}

Cells control the expression of genes in part through transcription factors, proteins which bind to particular sites along the genome and thereby enhance or inhibit the transcription of nearby genes (Fig \ref{f-scheme}). We can think of this transcriptional control process as an input/output device in which the input is the concentration of transcription factor and the output is the concentration of the gene product.  Although this qualitative picture has been with us for roughly forty years \cite{jacob+monod}, only recently have there been quantitative measurements of {\em in vivo} input/output relations and of the noise in output level when the input is fixed \cite{elowitz+al_02,ozbudak+al_02,blake+al_03,setty+al_03,raser+oshea_04,rosenfeld+al_05,pedraza+oudenaarden_05,golding+al_05,kuhlman+al_07,gregor+al_06b}. 
Because these input/output relations have a limited dynamic range, noise
limits the ``power'' of the cell to control gene expression levels. In this paper, we 
quantify these limits and derive the strategies that cells should use to take maximum advantage of the available power.  We show that, to make optimal use of its regulatory capacity, cells must achieve the proper quantitative matching among the input/output relation, the noise level, and the distribution of transcription factor concentrations used during the life of the cell.
We test these predictions against recent experiments on the
Bicoid and Hunchback  morphogens in the early {\em Drosophila} embryo \cite{gregor+al_06b}, and find that the observed distributions have a nontrivial structure which is in good agreement with theory, with no adjustable parameters.  This suggests that, in this system at least, cells make nearly optimal use of the available regulatory capacity and transmit substantially more than the simple on/off bit that might suffice to delineate a spatial expression boundary.

Gene expression levels ($g$) change in response to changes in transcription factor (TF)  concentration ($c$). These changes often are summarized by an input/output relation $\bar g(c)$ in which the mean expression level is plotted as a function of TF concentration (Fig \ref{f-scheme}). The average relationship is a smooth function but, because of noise, this does not mean that arbitrarily small changes in input transcription factor concentration are meaningful for the cell. The noise in expression levels could even be so large that reliable distinctions can only be made between (for example) ``gene on'' at high TF concentration and ``gene off'' at low TF concentration. To explore this issue, we need to quantify the number of reliably distinguishable regulatory settings of the transcription apparatus, a task to which Shannon's mutual information \cite{shannon48,cover+thomas} is ideally suited. While there are many ways to associate a scalar measure of correlation or control with a joint distribution of input and output signals, Shannon proved that mutual information is the only such quantity that satisfies certain plausible general requirements, independent of the details of the underlying distributions. Mutual information has been successfully used to analyze noise and coding in neural systems  \cite{spikes}, and it is natural to think that it may be useful for organizing our understanding of   gene regulation; see  also Ref \cite{ziv+al_06}.

Roughly speaking, the mutual information $I(c; g)$ between TF concentration and expression level counts the (logarithm of the) number of distinguishable expression levels achieved by varying $c$. If we measure the information in bits, then
\begin{equation}
I(c; g) = \int dc \, P_{TF} (c) \int dg \, P(g|c) \log_2 \left[
{{P(g|c)}\over {P_{\rm exp}(g)}} \right],
\label{mutI}
\end{equation}
where $P_{TF}(c)$ is the distribution of TF concentrations the cell generates in the course of its life, $P(g|c)$ is the distribution of expression levels at fixed $c$, and $P_{\rm exp}(g)$ is the resulting distribution of expression levels,
\begin{equation}
P_{\rm exp}(g) = \int dc \,  P(g|c) P_{TF}(c) .
\end{equation}
The distribution, $P(g|c)$, of expression levels at fixed transcription factor concentration describes the physics of the regulatory element itself, from the protein/DNA interaction, to the rates of protein synthesis and degradation; this distribution describes both the mean input/output relation {\em and} the noise  fluctuations around the mean output. The information transmission, or regulatory power, of the system is not determined by $P(g|c)$ alone, however, but also depends on the distribution, $P_{TF}(c)$, of transcription factor ``inputs''  that the cell uses, as can be seen from Eq (\ref{mutI}). By adjusting this distribution to match the properties of the regulatory element, the cell can maximize its regulatory power. 

Matching the distribution of inputs to the (stochastic) input/output relation of the system is a central concept in information theory \cite{cover+thomas}, and has been applied to the problems of coding in the nervous system. For sensory systems, the distribution of inputs is determined by the natural environment, and the neural circuitry can adapt, learn or evolve (on different times scales) to adjust its input/output relation. It has been suggested that maximizing information transmission is a principle which can predict the form of this adaptation \cite{barlow_61,laughlin_81,atick+redlich_90,brenner_00}. In transcriptional regulation, by contrast, it seems more appropriate to regard the input/output relation as fixed and ask how the cell might optimize its regulatory power by adjusting the distribution of TF inputs.

It is difficult to make  analytic progress in the general calculation of mutual information, but there is a simple and plausible approximation.
The expression level at a fixed TF concentration $c$ has a mean value $\bar g (c)$, which we can plot as an input/output relation (Fig \ref{f-scheme}). Let us assume that the fluctuations around this mean are Gaussian with a variance $\sigma_g^2 (c)$ which will itself depend on the TF concentration. Formally this means that
\begin{equation}
P(g|c) = {1\over{\sqrt{2\pi \sigma_g^2 (c)}}} \exp\Bigg{\{} - \, 
{{[g - \bar g (c)]^2}\over{2 \sigma_g^2 (c)}}  \Bigg{\}} . \label{gaussianio}
\end{equation}
Further let us assume that the noise level is small. Then we can expand all of the relevant integrals as a power series in the magnitude of $\sigma_g$: 
\begin{eqnarray}
I(c; g) &=& -\int d\bar g \,  \hat P_{\rm {exp}} (\bar g ) \log_2 \hat P_{\rm {exp}} (\bar g )\nonumber\\
&&\,\,\,\,\,
- {1\over 2} \int d\bar g \,  \hat P_{\rm {exp}} (\bar g ) \log_2 [2\pi e \sigma_g^2 (\bar g )] + \cdots ,
\label{inf_approx1}
\end{eqnarray}
where $\cdots$ are terms that vanish as the noise level decreases and $\hat P_{\rm {exp}} (\bar g )$ is the probability distribution for the average levels of expression. We can think of this as the distribution  that the cell is ``trying'' to generate, and would generate in the  absence of noise:
\begin{eqnarray}
\hat P_{\rm {exp}} (\bar g ) &\equiv& \int dc \,  P_{TF}(c) \delta [\bar g - \bar g (c)]\\
&=& P_{TF}(c = c_*(\bar g)){\Bigg |} {{d\bar g}\over {dc}} {\Bigg |}_{c = c_*(\bar{g})}^{-1},
\end{eqnarray}
where $c_* (\bar{g})$ is the TF concentration at which the mean expression level is $\bar{g}$;
similarly, by $\sigma_g({\bar g})$ we mean $\sigma_g (c)$ evaluated at $c = c_*(\bar{g})$. 

We now can ask how the cell should adjust these distributions to maximize the information being transmitted. In the low-noise approximation summarized by Eq (\ref{inf_approx1}), maximizing $I(c;g)$ poses a variational problem for $\hat P_{\rm exp}({\bar g})$ whose solution has a simple form:
\begin{eqnarray}
\hat P_{\rm {exp}}^* (\bar g ) &=& {1\over Z} \cdot {1\over {\sigma_g({\bar g})}}
\label{lownoise_Pout}\\
Z &=& \int d\bar g \, {1\over {\sigma_g({\bar g})}} .
\label{Z}
\end{eqnarray}
This result captures the intuition that effective regulation requires preferential use of signals that have high reliability or low variance---$\hat P_{\rm {exp}}^* (\bar g )$ is large where $\sigma_g$ is small. The actual information transmitted for this optimal distribution can be found by substituting $\hat P_{\rm {exp}}^* (\bar g )$ into Eq (\ref{inf_approx1}), with the result
$I_{\rm opt} (c;g) = \log_2 \left( {Z/{\sqrt{2\pi e}}}\right)$.

Although we initially formulated our problem as one of optimizing the distribution of {\em inputs}, the low noise approximation yields a result [Eq (\ref{lownoise_Pout})] which connects the optimal distribution of {\em output} expression levels to the {\em variances} of the same quantities, sampled across the life of a cell as it responds to natural variations in its environment. To the extent that the small noise approximation is applicable, data on the variance vs mean expression thus suffice to calculate the maximum information capacity;   details of the input/output relation, such as its degree of cooperativity, do not matter except insofar as they leave their signature on the noise.

Recent experiments  provide the data for an   application of these ideas. Elowitz and coworkers have measured  gene expression noise in a synthetic system, placing fluorescent proteins under the control of a lac--repressible promoter in {\em E. coli} \cite{elowitz+al_02}. Varying the concentration of an inducer, they determined the intrinsic variance of expression levels across a bacterial population as a function of mean expression level. Their results can be summarized as 
$\sigma_g^2({\bar g}) = a{\bar g} + b{\bar g}^2$,
where the expression level $g$ is normalized to have a maximum mean value of $1$,  and the constants are $a = 5 - 7 \times 10^{-4}$ and $b = 3 - 10 \times 10^{-3}$. Across most of the dynamic range ($\bar g \gg 0.03$), the small noise approximation should be valid and, as discussed above, knowledge of $\sigma_g(\bar g )$ alone suffices to compute the optimal information transmission. We
find $I_{\rm opt}(c;g) \sim 3.5~$bits: rather than being limited to on/off switching, these  transcriptional control systems could in principle specify $2^{I_{\rm opt}} \sim 10 - 12$ distinguishable levels of gene expression! 
It is not clear whether this capacity, measured in an engineered system, is available to or used by
{\em E. coli}  in its natural environment. The calculation does demonstrate, however, that optimal information transmission values  derived from real data are 
more than one bit, but perhaps small enough to provide significant constraints on regulatory function.

When the noise is not small, no simple analytic approaches are available. On the other hand, so long as $P(g|c)$ is known explicitly, our problem is equivalent to one well--studied in communication theory, and efficient numerical algorithms are available for finding the input distribution $P_{TF}(c)$ that optimizes the information $I(c;g)$ defined in Eq (\ref{mutI}) \cite{blahut}. In general we must extract $P(g|c)$ from experiment and, to deal with finite data, we will assume that it has the Gaussian form of Eq (\ref{gaussianio}). $P(g|c)$  then is completely determined by measuring just two functions of $c$: the mean input/output relation $\bar g (c)$ and the output variance  $\sigma_g^2 (c)$. The central point is that, in the general case, solving the information optimization problem requires {\em only} empirical data on the input/output relation and noise.

\begin{figure} 
   \centering
 \includegraphics[width=\linewidth]{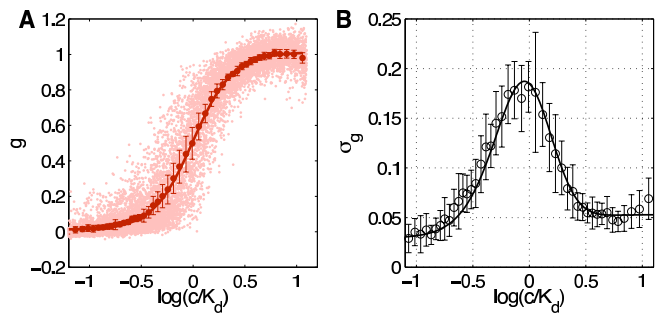}
   \caption{The Bcd/Hb input/output relationship 
   in the {\em Drosophila melanogaster} syncitium at early nuclear cycle 14 \cite{gregor+al_06b}. (a) Each point marks the Hb ($g$) and Bcd ($c$) concentration in a single nucleus, as inferred from immunofluorescent staining;  data are from $\sim 11\E{3}$ individual nuclei across 9 embryos.  Hb expression levels $g$ are  normalized so that the  maximum and minimum mean expression levels are 1 and 0 respectively; small errors in the estimate of  background fluorescence result in some apparent expression values being slightly negative. Bcd concentrations $c$ are normalized by $K_d$,  the concentration of Bcd at which the mean Hb expression level is  half maximal. For details of normalization across embryos, see \cite{gregor+al_06b}.  Solid red line is a sigmoidal fit to the mean $g$ at each value of $c$, and error bars are $\pm$ one s.e.m.. (b) Noise in Hb as a function of Bcd concentration; error bars are $\pm$ one s.d. across embryos, and the curve is a  fit  from Ref \cite{tkacik+al_07}.}
\label{dros_inout}
\end{figure}

The initial events of pattern formation in the embryo of the fruit fly {\em Drosophila} provide a promising testing ground for the optimization principle proposed here.
These events  depend on the establishment of spatial gradients in the concentration of various morphogen molecules, most of which are transcription factors \cite{wolpert_69,lawrence_92}. To be specific, consider the response of the {\em hunchback} (Hb) gene to the maternally established gradient of the transcription factor Bicoid (Bcd)  \cite{driever+nusslein-volhard_88a,driever+nusslein-volhard_88b,driever+nusslein-volhard_89,struhl+al_89}. A recent experiment reports the Bcd and Hb concentrations in thousands of individual nuclei of the {\em Drosophila} embryo, using fluorescent antibody staining \cite{gregor+al_06b}; the results can be summarized by the mean input/output relation and noise level shown in Fig \ref{dros_inout}. These data  can be understood in some detail on the basis of a simple physical model  \cite{tkacik+al_07}, but here we   use the experimental observations directly to make phenomenological predictions about maximum available regulatory power and optimal distribution of expression levels. 

\begin{figure}[bht] 
   \centering
 \includegraphics[width=0.8\linewidth]{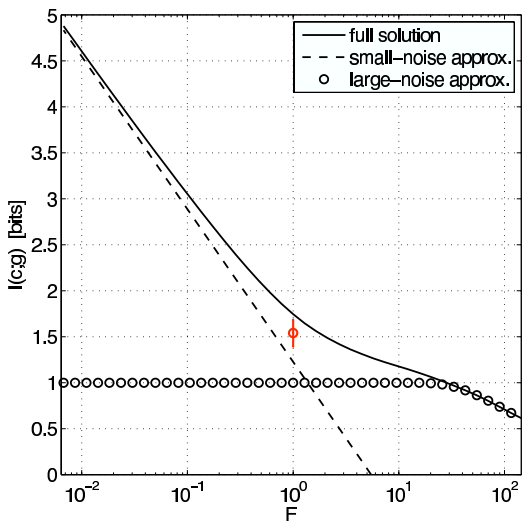}
   \caption{Optimal information transmission for the Bcd/Hb system as a function of the noise variance rescaling factor $F$. $1/F$ is approximately equal to the factor by which the number of input and output signaling molecules has to be increased for the corresponding gain in capacity. Dashed and dotted curves show the solutions in the small-noise and large-noise approximations, respectively. The real system, $F=1$, lies in an intermediate region where neither the small-noise nor the large-noise approximation are valid. Measured information $I_{\rm data}(c;g)$ shown in red (errorbar is s.d. over 9 embryos).} 
\label{f-axs}
\end{figure}

Given the measurements of the mean input/output relation $\bar g (c)$ and noise $\sigma_g(c)$ shown in Fig \ref{dros_inout}, we can calculate the maximum mutual information between Bcd and Hb concentrations by following the steps outlined above; we find $I_{\rm opt}(c;g) = 1.7\,{\rm bits}$.  To place this result in context, we imagine a system that has the same mean input/output relation, but the noise variance is scaled by a factor $F$, and ask how the optimal information transmission depends on $F$.  This is not just a mathematical trick:  for most physical sources of noise, the relative variance is inversely proportional to the number of molecules, and so scaling the expression noise variance down by a factor of ten is equivalent to assuming that all relevant molecules are present in ten times as many copies.  We see in Fig \ref{f-axs} that there is a large regime in which the regulatory power is well approximated by the small noise approximation. In the opposite extreme, at large noise levels, we expect that there are (at best!) only two distinguishable states of high and low expression, so that our problem approaches the asymmetric binary channel \cite{Silverman_55}.
The exact result interpolates smoothly between these two limiting cases with the real system ($F=1$) lying closer to the small noise limit, but deviating from it significantly. 

In the  embryo, maximizing  information flow from transcription factor to target gene has a very special meaning. 
Cells acquire ``positional information,'' and thus can take actions which are appropriate to their position in the embryo, by responding to the local concentration of morphogen molecules \cite{wolpert_69}.
In the original discussions, ``information'' was used colloquially. 
But in the simplest picture of   {\em Drosophila} development \cite{lawrence_92,rivera-pomar+jackle_96}, information in the technical sense really does flow from physical position along the anterior--posterior axis to the concentration of the primary maternal gradients (such as Bcd) to the expression level of the gap genes (such as Hb).  Maximizing the mutual information between Bcd and Hb  thus  maximizes the positional information that can be carried by the Hb expression level.   

More generally, rather than thinking of each gap gene as having its own spatial profile, we can think of the expression levels of all the gap genes together as a code for the position of each cell. 
In the same way that  the four bases (two bits) of DNA must code in triplets in order to represent arbitrary sequences of 20 amino acids, we can ask how many gap genes would be required to encode a unique position in the $N_{\rm rows} \sim 100$ rows of nuclei along the anterior--posterior axis.
If the regulation of Hb by Bcd is typical of what happens at this level of the developmental cascade, then each letter of the code is limited to less than two bits ($I_{\rm opt} = 1.7\,{\rm bits}$) of precision;  since $\log_2(N_{\rm rows})/I_{\rm opt} = 3.9$,  the code would need to have at least four letters.  It is interesting, then, to note that there are four known gap genes---{\em hunchback, kr\"uppel, giant} and {\em knirps} \cite{rivera-pomar+jackle_96}---which provide the initial readout of the maternal anterior--posterior gradients.

\begin{figure}[b] 
   \centering
 \includegraphics[width=0.9\linewidth]{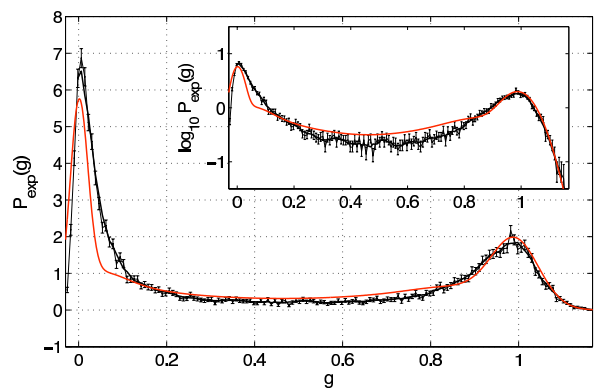}
   \caption{The measured (black) and optimal (red) distributions of Hunchback expression levels. The measured distribution is estimated from data  of Ref \cite{gregor+al_06b}, by making  a histogram of the $g$ values for each data point in Fig \ref{dros_inout}. The optimal solution corresponds to the capacity of $I_{\rm opt}(c;g) = 1.7\,{\rm bits}$. The same plot is shown on logarithmic scale in the inset.} 
   \label{f-droso}
   \end{figure}

Instead of plotting Hunchback expression levels vs either position or Bcd concentration, we can ask about the {\em distribution} of expression levels seen across all nuclei,  $P_{\rm exp}(g)$, as shown in Fig \ref{f-droso}.
The  distribution  is bimodal, so that large numbers of nuclei have near zero or near maximal  Hb, consistent with the idea that there is an expression boundary---cells in the anterior of the emrbyo have Hb ``on'' and cells in the posterior  have Hb ``off.''  
But intermediate levels of Hunchback expression also occur with nonzero probability, and the overall distribution is quite smooth.  We can compare this experimentally measured distribution with the distribution predicted if the system maximizes information flow, and we see from Fig \ref{f-droso} that the agreement is quite good.  
The optimal distribution reproduces the bimodality of the real system, hinting in the direction of a simple on/off switch, but also correctly predicts that the system makes use of intermediate expression levels.  From the data we can also compute directly the mutual information between Bcd and Hb levels, and we find $I_{\rm data}(c;g)=1.5 \pm 0.15\,{\rm bit}$, or
$\sim 90\%$ ($0.88\pm 0.09$) of the theoretical maximum.

The agreement between the predicted and observed distributions of Hunchback expression levels is encouraging.  We note, however, some caveats.    Bicoid has multiple targets and many of these genes have multiple inputs \cite{Espinosa_05}, so to fully optimize information flow we need to think about a more complex problem than the single input, single output system considered here.  Measurement of the distribution of expression levels requires a fair sampling of all the nuclei in the embryo, and this was not the intent of the experiments of \cite{gregor+al_06b}.  Similarly, the theoretical predictions depend somewhat on the behavior of the input/output relation and noise at low expression levels, which are difficult to characterize experimentally, as well as the (possible) deviations from Gaussian noise. A complete test of our theoretical predictions will thus require a new generation of experiments.  

In summary, the functionality of a transcriptional regulatory element is determined by a combination of its input/output relation, the noise level, and the dynamic range of transcription factor concentrations used by the cell.  In parallel to discussions of neural coding \cite{laughlin_81,brenner_00}, we have suggested that organisms can make maximal use of the available regulatory power by achieving consistency among these three different ingredients; in particular, if we view the input/output relation and noise level as fixed, then the distribution of transcription factor concentrations or expression levels is predicted by the optimization principle.  Although many aspects of transcriptional regulation are well studied, especially in unicellular organisms, these distributions of protein concentrations have not been investigated systematically. In embryonic development, by contrast, the distributions of expression levels can literally be read out from the spatial gradients in morphogen concentration. We have focused on the simplest possible picture, in which a single input transcription factor regulates a single target gene, but nonetheless find encouraging agreement between the predictions of our optimization principle and the observed distribution of the Hunchback morphogen in {\em Drosophila}. We emphasize that our prediction  is not the result of a model with many parameters; instead we have a theoretical principle for what the system ought to do so as to maximize its performance, and no free parameters.

\vskip - 0.075in
{\small We thank T Gregor, DW Tank \& EF Wieschaus for many helpful discussions, as well as for sharing the raw data from Ref \cite{gregor+al_06b}. This work was supported in part by NIH grants P50 GM071508 and R01 GM077599,  by the Burroughs Wellcome Fund Program in Biological Dynamics (GT) and  by US Department of Energy grant DE-FG02-91ER40671 (CC).}
\end{document}